\documentclass[aps,prl,twocolumn,showkeys,superscriptaddress]{revtex4-1}

\usepackage[artemisia]{textgreek}
\usepackage{amssymb}
\usepackage{epsfig}
\usepackage{amsmath}
\usepackage{graphicx}
\usepackage{dcolumn}
\usepackage{latexsym}
\usepackage{color}
\usepackage{epstopdf}
\usepackage{subfigure}
\usepackage{comment}
\usepackage{nicefrac}
\usepackage{blindtext}
\usepackage[ulem=normalem]{changes}
\usepackage{float}
\usepackage{xcolor}
\usepackage{soul}
\usepackage[T1]{fontenc}
\usepackage[hidelinks]{hyperref}
\usepackage{xcolor}
\setcitestyle{super}
\hypersetup{
    colorlinks,
    linkcolor={red!50!black},
    citecolor={blue!50!black},
    urlcolor={blue!80!black}
}

\newcommand*{\citen}[1]{%
  \begingroup
    \romannumeral-`\x 
    \setcitestyle{numbers}%
    \cite{#1}%
  \endgroup   
}

\begin{document}
\title{Field effect enhancement in buffered quantum nanowire networks}

\author{Filip Krizek\textsuperscript{\textdagger}}
\affiliation{Center For Quantum Devices and Station Q Copenhagen, Niels Bohr Institute, University of Copenhagen, 2100 Copenhagen, Denmark}
\altaffiliation{These authors contributed equally to this work.}

\author{Joachim E. Sestoft\textsuperscript{\textdagger}}
\affiliation{Center For Quantum Devices and Station Q Copenhagen, Niels Bohr Institute, University of Copenhagen, 2100 Copenhagen, Denmark}
\altaffiliation{These authors contributed equally to this work.}

\author{\mbox{Pavel Aseev\textsuperscript{\textdagger}}}
\affiliation{QuTech and Kavli Institute of Nanoscience, Delft University of Technology, 2600 GA Delft, The Netherlands}
\altaffiliation{These authors contributed equally to this work.}

\author{Sara Marti-Sanchez}
\affiliation{Catalan Institute of Nanoscience and Nanotechnology (ICN2), CSIC and BIST, Campus UAB, Bellaterra, Barcelona, Catalonia, Spain}

\author{Saulius Vaitiek\.{e}nas}
\affiliation{Center For Quantum Devices and Station Q Copenhagen, Niels Bohr Institute, University of Copenhagen, 2100 Copenhagen, Denmark}

\author{Lucas Casparis}
\affiliation{Center For Quantum Devices and Station Q Copenhagen, Niels Bohr Institute, University of Copenhagen, 2100 Copenhagen, Denmark}

\author{Sabbir A. Khan}
\affiliation{Center For Quantum Devices and Station Q Copenhagen, Niels Bohr Institute, University of Copenhagen, 2100 Copenhagen, Denmark}

\author{Yu Liu}
\affiliation{Center For Quantum Devices and Station Q Copenhagen, Niels Bohr Institute, University of Copenhagen, 2100 Copenhagen, Denmark}

\author{Toma\v{s} Stankevi\v{c}}
\affiliation{Center For Quantum Devices and Station Q Copenhagen, Niels Bohr Institute, University of Copenhagen, 2100 Copenhagen, Denmark}

\author{Alexander M. Whiticar}
\affiliation{Center For Quantum Devices and Station Q Copenhagen, Niels Bohr Institute, University of Copenhagen, 2100 Copenhagen, Denmark}

\author{Alexandra Fursina}
\affiliation{Microsoft Station Q, Delft University of Technology, 2600 GA Delft, The Netherlands}

\author{Frenk Boekhout}
\affiliation{QuTech and Netherlands Organization for Applied Scientific Research (TNO), Stieltjesweg 1, 2628 CK Delft, The Netherlands}

\author{Rene Koops}
\affiliation{QuTech and Netherlands Organization for Applied Scientific Research (TNO), Stieltjesweg 1, 2628 CK Delft, The Netherlands}

\author{Emanuele Uccelli}
\affiliation{QuTech and Netherlands Organization for Applied Scientific Research (TNO), Stieltjesweg 1, 2628 CK Delft, The Netherlands}

\author{Leo P. Kouwenhoven}
\affiliation{QuTech and Kavli Institute of Nanoscience, Delft University of Technology, 2600 GA Delft, The Netherlands}
\affiliation{Microsoft Station Q, Delft University of Technology, 2600 GA Delft, The Netherlands}

\author{Charles M. Marcus}
\affiliation{Center For Quantum Devices and Station Q Copenhagen, Niels Bohr Institute, University of Copenhagen, 2100 Copenhagen, Denmark}

\author{Jordi Arbiol}
\affiliation{Catalan Institute of Nanoscience and Nanotechnology (ICN2), CSIC and BIST, Campus UAB, Bellaterra, Barcelona, Catalonia, Spain}
\affiliation{ICREA, Pg. Llu\'{i}s Companys 23, 08010 Barcelona, Catalonia, Spain}

\author{Peter Krogstrup}
\email{krogstrup@nbi.dk}
\affiliation{Center For Quantum Devices and Station Q Copenhagen, Niels Bohr Institute, University of Copenhagen, 2100 Copenhagen, Denmark}

\date{\today}

\begin{abstract}
 
III-V semiconductor nanowires have shown great potential in various quantum transport experiments. However, realizing a scalable high-quality nanowire-based platform that could lead to quantum information applications has been challenging. Here, we study the potential of selective area growth by molecular beam epitaxy of InAs nanowire networks grown on GaAs-based buffer layers. The buffered geometry allows for substantial elastic strain relaxation and a strong enhancement of field effect mobility. We show that the networks possess strong spin-orbit interaction and long phase coherence lengths with a temperature dependence indicating ballistic transport. With these findings, and the compatibility of the growth method with hybrid epitaxy, we conclude that the material platform fulfills the requirements for a wide range of quantum experiments and applications.

\end{abstract}
\pacs{}
\maketitle

Material science plays a key role in quantum computing research. Long quantum state lifetimes \--- the fundamental prerequisite for realizing quantum computers \--- rely on the ability to produce materials with high purity and structural quality. Together with the requirements of scalability and reproducibility, these properties are what mainly defines the challenges of material science in quantum computing today. Proposals for topological quantum computing,\cite{oreg2010helical,lutchyn2010majorana, sarma2015majorana} which are based on hybrid semiconductor-superconductor nanowire (NW) networks, are being pursued by numerous research groups and have ignited intense research efforts on hybrid epitaxy. \cite{krogstrup2015epitaxy, Shabani2016, kang2017robust,sestoft2017hybrid,gazibegovic2017epitaxy} NW scalability is tightly related to the semiconductor growth approach. Top-down lithography has been used to define NWs in two-dimensional layers\cite{Shabani2016, Nichele2017} and a variety of methods have been pursued for alignment and positioning of bottom-up vapor-liquid-solid (VLS) grown NWs, such as dielectrophoresis techniques,\cite{freer2010high} nanoscale combing\cite{yao2013nanoscale} and magnetic aligning of NWs.\cite{hangarter2005magnetic} Despite of these developments, large-scale synthesis of bottom-up grown high-mobility NW networks that are compatible with epitaxial interwire connections and semiconductor/superconductor epitaxy has still not been realized. To realize the epitaxial connections, a lot of effort has been put into the growth of branched NWs via the VLS method.\cite{kang2013crystal,car2014rationally,gazibegovic2017epitaxy,krizek2017growth} A scalable approach has been developed in Ref. \citen{borg2015mechanisms, schmid2015template} using template assisted growth of in-plane NW networks.\cite{gooth2017ballistic} Nonetheless, this approach is not yet compatible with superconductor epitaxy. An alternative scalable approach is to use lithographically defined openings in a mask on a crystalline substrate. This method is referred to as selective area growth (SAG) and until recently has mainly been used in conjunction with metal organic chemical vapour deposition \cite{joyce1962mocvd, shaw1966selective}, metal organic vapour phase epitaxy\cite{kato1994selective,mohan2003fabrication}, chemical beam epitaxy and metal organic molecular beam epitaxy (chemical beam epitaxy).\cite{vodjdani1982parametric, tokumitsu1984molecular, andrews1989selective, furuhata1991selective} In contrast to molecular beam epitaxy (MBE), the dissociation kinetics of the chemical precursors in these methods enhance the growth selectivity on masked substrates by expanding the growth parameter window, but typically at a cost of crystal purity. Even though the initial work was reported about 30 years ago,\cite{yokoyama1989low, okamoto1987selective, sugaya1992selective, kuroda1993selective} only the recent promising results reported in Ref. \citen{desplanque2014inas, desplanque2014influence, tutuncuoglu2015towards, fahed2016selective, Anna2018} have renewed the interest in SAG by MBE.

\begin{figure*}[ht!]
\vspace{0.2cm}
\includegraphics[scale=0.25]{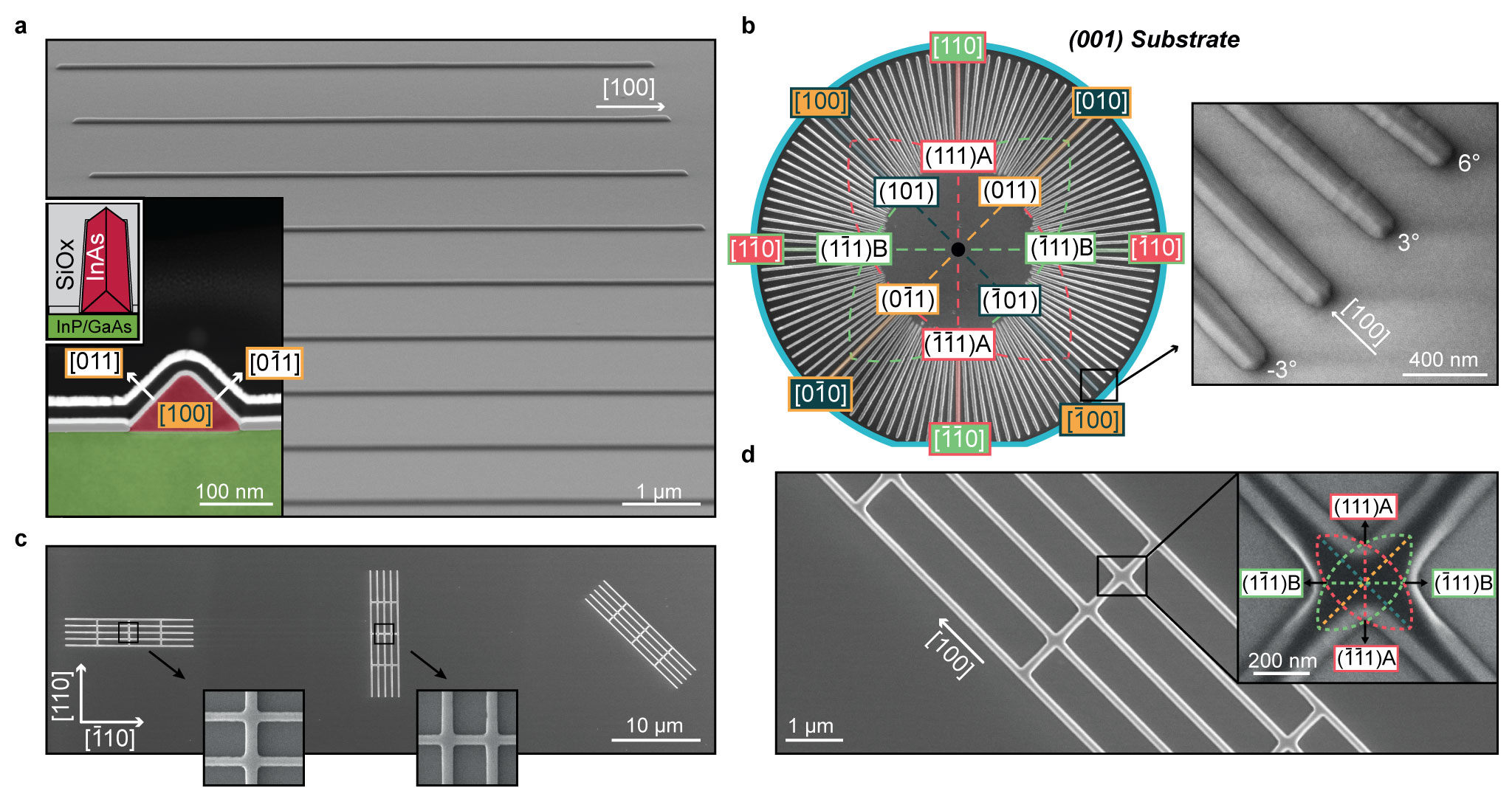}
\vspace{0.2cm}
\caption{\textbf{Selective area grown InAs NW networks.} \textbf{a}, SEM micrograph of 10 \textmu m long [100] InAs NWs grown on InP (001) with a  SiO\textsubscript{x} mask. Inset shows a HAADF-STEM cross-section image of a NW, where the orientation and faceting is color-correlated to the stereographic projection in \textbf{b}. \textbf{b}, Stereographic projection including the high-symmetry orientations of a [100] substrate, where the perimeter corresponds to in-plane NW directions (box fill color), with the facets normal to the NW given by the perpendicular orientations (corresponding border color). The graphics overlay an SEM micrograph of NWs selectively grown at increments of 3$^{\circ}$. Zoom-in shows a [100] NW grown next to off-axis NWs with roughened and vicinal faceting. \textbf{c}, SEM micrographs of three NW networks aligned along the [110], [1$\bar{1}$0] and [100] directions. \textbf{d}, SEM micrograph of [100]/[010] NW network junctions. Zoom-in shows how the crystal symmetry results in two-fold symmetric junctions indicated by the polar (111)A and (111)B facets.}
\label{fig1}
\end{figure*}

In this work, we present selective area growth of InAs NW networks by MBE, which are grown either on GaAs based buffer layers or directly on semi-insulating InP and GaAs substrates. We demonstrate growth of lithographically designed NW networks with well-defined junctions, where the faceting depends on the mask alignment to the crystal orientation of the substrate. We selectively grow Sb-dilute GaAs buffer layers with flat top-facets that protrude out of the substrate plane and allow for significant elastic strain relaxation of the InAs. The improved interface quality results in enhanced field effect response close to conductance pinch-off. In addition, magneto-conductance experiments show strong spin-orbit coupling and phase coherence where the temperature dependence indicates ballistic transport. Therefore, the compatibility of the SAG NW platform with the growth of
epitaxial superconductors on selected facets\cite{krogstrup2015epitaxy} demonstrates its potential for large-scale applications in the field of gateable superconducting electronics (see Supplementary Information S1 and Ref. \citen{sole2018sag} for details).

\section{Results and Discussion}

The NWs are grown on semi-insulating InP and GaAs substrates covered with thin dielectric masks, at relatively low growth rates and at growth temperatures similar to those used for corresponding planar growth by MBE, see \hyperlink{foo}{Methods}. The mask openings were defined by standard electron beam lithography (EBL) patterning and selective etching, as described in Supplementary Information S2. The substrates become fully insulating at low temperatures and are therefore suitable for as-grown device fabrication and transport experiments directly on the growth substrate. The scanning electron microscope (SEM) micrograph in Fig. \ref{fig1}\textbf{a} shows an array of InAs NWs oriented along the [100] direction on a (001) Fe-doped InP substrate. For SAG of InAs NWs on GaAs substrates, see Supplementary Information S3. The NWs exhibit smooth (011) and (0$\bar{1}$1) facets, as shown in the colored high-angle annular dark-field scanning transmission electron microscope (HAADF-STEM) image in the Fig. \ref{fig1}\textbf{a} inset. 
The uniformity of individual NW facets depends on the growth conditions \cite{desplanque2014influence, fahed2016selective} and the quality of the pre-patterned substrate.~Moreover, uniform, high-symmetry facets were found only on NWs oriented along the high-symmetry [100], [010], [1$\bar{1}$0] and [110] crystal directions, as illustrated in the stereographic projection in Fig. \ref{fig1}\textbf{b}. For instance, a NW oriented along the [$\bar{1}$00] direction has \{0$\bar{1}$1\}, family facets due to local cusps in surface energy. Even though the roughness of the NWs depends on growth conditions, a slight misalignment with respect to the high-symmetry crystal orientation causes vicinal faceting, as shown in the zoom-in of Fig. \ref{fig1}\textbf{b}. Consequently, there are constraints on the in-plane directions, which set the overall symmetry and design of the NW networks. In the case of networks grown on (001) substrates, there are eight high-symmetry in-plane directions (indicated on the perimeter of the stereographic projection). As a result, one junction can be connected to eight NWs. There are two families of networks on (001) substrates consisting of perpendicularly oriented NWs, the <110>/<1$\bar{1}$0> type and the <100>/<010> type (see Fig. \ref{fig1}\textbf{c}). At given growth conditions, both types of junctions exhibit a four-fold symmetric morphology for short growth times. However, as the NWs grow thicker, the junctions tend to become two-fold symmetric, see Fig. \ref{fig1}\textbf{d}. In the case of the <110>/<1$\bar{1}$0> junction the symmetry breaking is related to the growth kinetics, where the difference in adatom diffusion lengths along the [110] and [1$\bar{1}$0] directions causes the material to incorporate easier along the [110] direction.\cite{krogstrup2013advances} At the given growth time and conditions, the <110>/<1$\bar{1}$0> NW junctions maintain a non-tapered four-fold-like morphology as seen in the inset of Fig. \ref{fig1}\textbf{c}. In the <100>/<010> case, the symmetry breaking appears earlier, likely due to the different polarity of faceting of the two orientations, as indicated in Fig. \ref{fig1}\textbf{d}.

\begin{figure}[t!]
\vspace{0.2cm}
\includegraphics[scale=1]{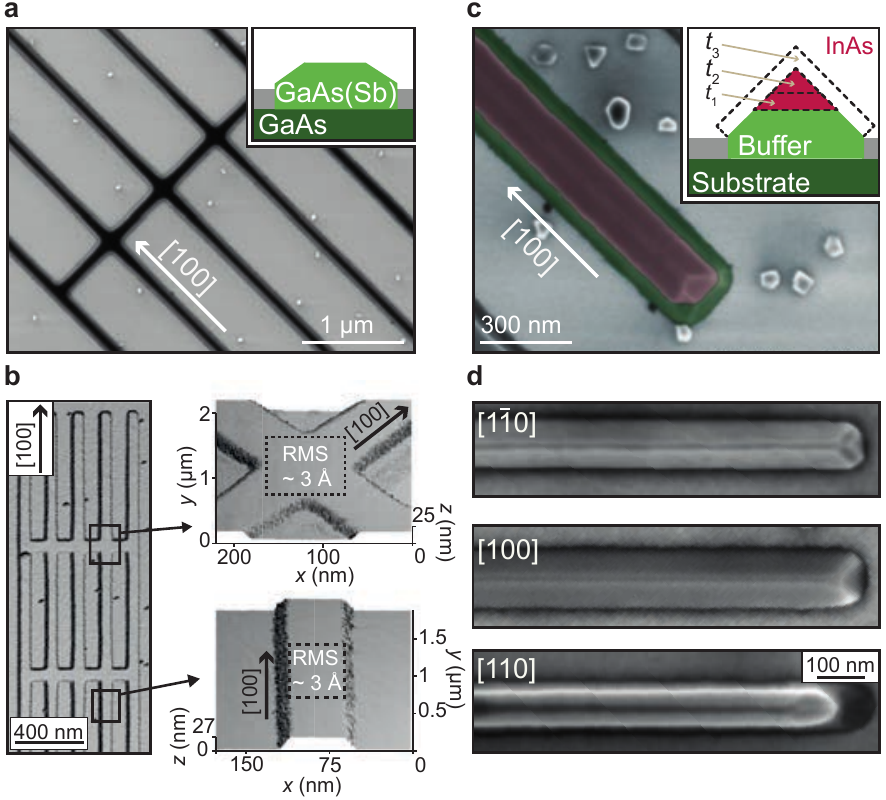}
\vspace{0.2cm}
\caption{\textbf{Selectively grown GaAs(Sb) buffer.} \textbf{a}, SEM micrograph of a selectively grown GaAs(Sb) NW network. \textbf{b}, AFM scans showing flat (001) top-facets of GaAs(Sb) networks aligned along the [100]/[010] directions. Zoom-ins on highlighted areas show the extracted roughness of the top-facet of a junction and a straight NW. \textbf{c}, SEM micrograph of InAs NWs selectively grown on the top-facet of the buffer. Inset shows the InAs NW morphology evolution with growth time, $t_{1-3}$.~\textbf{d}, SEM micrographs of three types of high-symmetry oriented InAs NWs grown on the buffer.} 
\label{fig2}
\end{figure}

We find that top-gated InAs NWs grown directly on non-buffered InP or GaAs substrates generally display a weak field effect response close to the conductance pinch-off region. This is most likely related to the NW/substrate interface. To enhance the electrical properties of the NWs we focus on improving the quality of the interface and turn our attention to GaAs\textsubscript{1-x}Sb\textsubscript{x} buffers where the lattice matching can be tuned from GaAs to InAs by changing the composition, x. 


\begin{figure*}[ht!]
\vspace{0.2cm}
\includegraphics[scale=1]{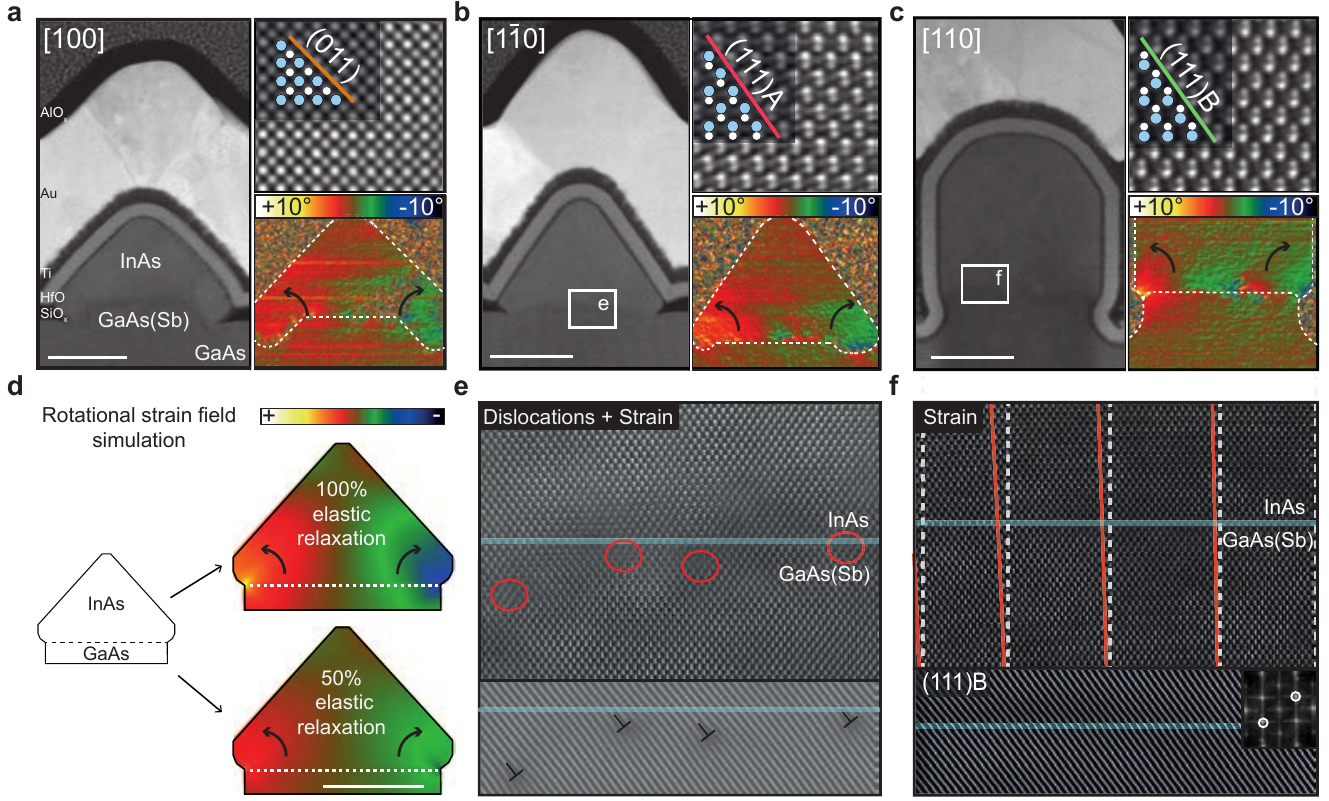}
\vspace{0.2cm}
\caption{\textbf{Elastic strain relaxation of InAs on GaAs(Sb) buffer.} \textbf{a-c}, Cross-sectional TEM micrographs of top-gated devices with InAs NWs grown on top of GaAs(Sb) buffer layers along the [100], [1$\bar{1}$0] and [110] orientations. Top right insets are cross-sectional HAADF-STEM images of the InAs NW segments with pure crystalline structures, where the illustrated high-symmetry plane highlights the related facet polarity (As atoms: white, In atoms: blue). Bottom right insets show GPA rotational maps around the (111) planes. Scale-bars are \mbox{50 nm}. \mbox{\textbf{d}, GPA} rotational map simulation for a [100] oriented buffered NW showing elastic strain relaxation profiles. Scale-bar is \mbox{50 nm}. \textbf{e}, Atomic resolution aberration corrected HAADF-STEM  micrograph of the interface indicated in \textbf{b} showing misfit dislocations and strain. The bottom panel is the same image after FFT filtering, with a highlighted aperiodic misfit dislocation array. The blue line indicates the interface. \textbf{f}, Atomic resolution aberration corrected HAADF-STEM  micrograph of the interface indicated in \textbf{c} showing no misfit dislocations and pronounced strain as indicated by the white dashed lines (no strain) and red line (the actual plane displacement). The bottom panel is the same image after FFT filtering, with no visible misfit dislocations. }
\label{fig3}
\end{figure*}

In Fig. \ref{fig2}\textbf{a} we show an SEM micrograph of a GaAs(Sb) buffer NW network grown on a semi-insulating (001) GaAs substrate at typical planar GaAs growth temperatures, see \hyperlink{foo}{Methods}. We find that the buffer layer has flat (001) top-facets, with a root-mean-square (RMS) roughness of $\sim$ 3 \AA{}. The roughness is uniform across the whole wafer as shown on both single NWs and within the junctions in Fig. \ref{fig2}\textbf{b} (see the Supplementary Information S4 for analysis of the other NW orientations). We emphasize that the flatness of the buffer is a crucial step towards obtaining a low disorder interface to the InAs transport channel. Compositional analysis performed by electron energy loss spectroscopy (EELS) and energy dispersive x-ray spectroscopy (EDX) reveals a very low fraction of Sb (close to the detection limit of $\sim$2\%). This is lower than the calibrated flux ratios, which would predict $\sim$7\% if the incorporation rates of As and Sb were equal. The low incorporation efficiency of Sb is consistent with previous reports.\cite{ahmad2017two,sarney2014influence} Despite the low Sb incorporation, the surfactant properties of Sb could play an important role in the growth kinetics, and potentially promote a more smooth layer by layer buffer growth. The details of this effect are to be further investigated in forthcoming experiments.

In Fig. \ref{fig2}\textbf{c} we show InAs selectively grown solely on the top-facet of the GaAs(Sb) buffer. The InAs was grown in the lower temperature bound of the selective growth window, which is apparent by the crystallites on the oxide mask. The morphology of the NWs grown on the buffer is highly dependent on the growth time and the width of the buffer top-facet. In the inset of Fig. \ref{fig2}\textbf{c} we show an illustration of how a predicted evolution of the growth would look like if it was thermodynamically driven, i.e. able to reach the lowest free energy (or equilibrium shape) for any given volume. The equilibrium shape of a crystal results from minimizing its anisotropic surface free energy under the constraint of constant volume. If there are additional constraints, such as a mask opening into which the crystal is confined to, then the equilibrium shape will depend on its volume. Assuming that the cross-sectional shape of the SAG NWs are equilibrium shapes, the NWs will most likely first grow solely on the top-facets of the buffer until a fully faceted shape is reached, as illustrated in the inset of Fig. \ref{fig2}\textbf{c} at growth time $t_2$. This particular equilibrium shape is also of special interest for realization of high quality SAG NWs, for reasons discussed below. It is clear, that the ratio of the buffer/NW growth time affects the shape and dimensions of the NW structures, and understanding the detailed processes whether it is in thermodynamical or kinetically driven regimes will be subject of ongoing studies.


The top-view SEM micrographs in Fig. \ref{fig2}\textbf{d} show the morphology of the [1$\bar{1}$0], [100] and [110] high-symmetry NWs, where the InAs NWs completely covers the buffer layer. Interestingly, the incorporation rate is higher on the [110] NWs in this growth. The reasons can be of both thermodynamic or kinetic origin, i.e. due to lower surface energies or lower activation barriers for incorporation, respectively. The latter is a viable possibility, since adatom kinetics on (001) surfaces is known to be highly anisotropic.\cite{ohta1989anisotropic}

Control over the width of the mask opening becomes difficult for features below 90 nm when wet etching is used. However, implementation of the buffer layer, where the top-facet width is decreasing with growth time, also provides an in-situ method for tuning the InAs NW width. This opens up for the possibility of engineering thinner NWs that are not in contact with either the oxide mask or the processed mask-opening (see Supplementary Information S5).

Atomic resolution and aberration corrected HAADF-STEM characterization was performed on cross-sectional cuts of top-gated NW devices in the three high-symmetry orientations. Figure \ref{fig3} shows the distinct cross-sectional shapes of \textbf{a} [100], \textbf{b} [1$\bar{1}$0] and \textbf{c} [110] InAs NWs, which were grown on the GaAs(Sb) buffer. The HAADF-STEM micrographs show that there are no threading dislocations running through the NWs (as also reported in Ref. \citen{fahed2016selective}) and that the bulk structure is single crystalline for all the three NW orientations. The highlighted facet planes correspond to the stereographic projection shown in Fig.~\ref{fig1}\textbf{b} which will also determine the corresponding NW shape. We further note that the three different types of facets, non-polar, A-polar and B-polar, are likely to have different electron affinities and provide additional band alignment options when optimizing contact to superconductors or metals.

The relative lattice mismatch between the InAs NW and the InP or GaAs substrates is 3\% and 7\%, respectively.~As shown in Supplementary Information S6, when growing directly on InP substrates, without a buffer layer, the lattice mismatch to InAs is fully compensated by relaxation at the interface via periodic arrays of misfit dislocations. That results in an abrupt change of the lattice spacing at the substrate/NW interface. The situation is different for the buffered NWs, where the buffer protrudes out of the substrate plane. Consequently, the InAs NWs have significantly more freedom to make the elastic rotation needed to relax to its equilibrium lattice parameter without introducing dislocations. This effect is observed on all analyzed buffered GaAs(Sb)//InAs interfaces, which is apparent from the gradual change in lattice constant and significantly lower density of misfit dislocations.

 \begin{figure}[t!]
 \vspace{0.2cm}
 \includegraphics[scale=1]{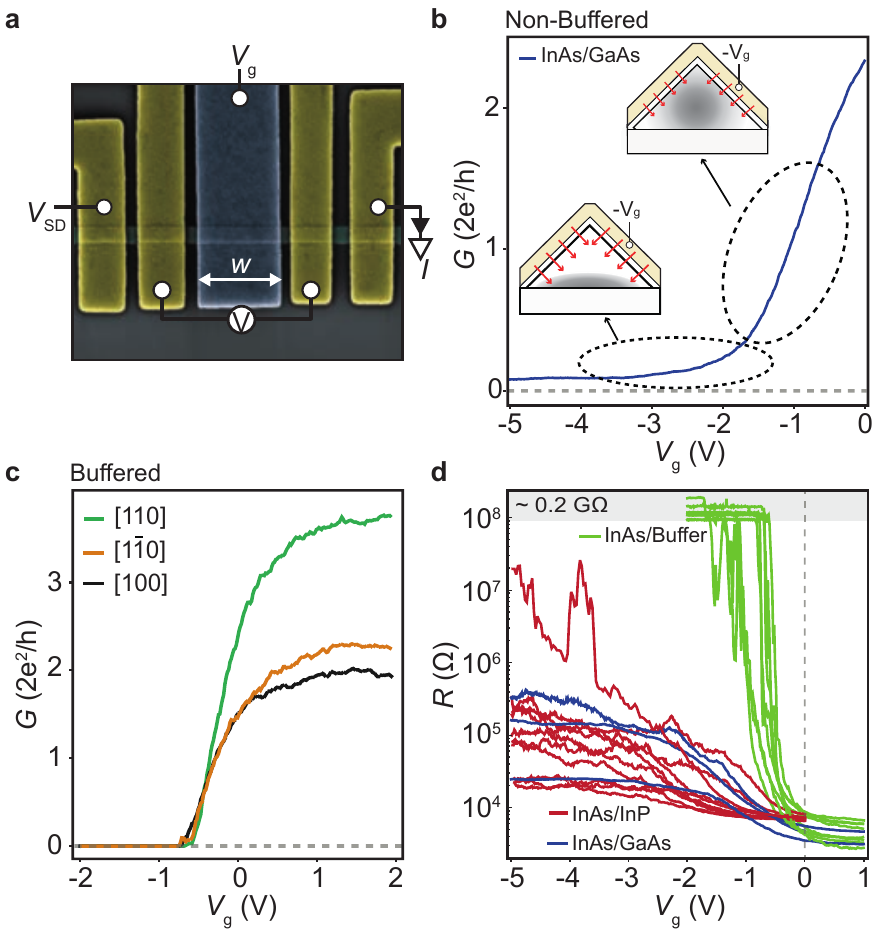}
 \vspace{0.2cm}
 \caption{\textbf{Carrier density tuning.} \textbf{a}, False-colored SEM micrograph of a typical top-gated four-probe NW device. Yellow, Ti/Au contacts; blue, gates; grey, InAs NW; $V$\textsubscript{SD}, bias voltage; $I$, measured current; $V$\textsubscript{g}, gate voltage controlling the chemical potential; $W$, gate width; \textbf{b}, Conductance, $G$, as a function of $V$\textsubscript{g}, for InAs grown on a GaAs substrate. Sketches illustrate the carrier density distribution as a function of $V$\textsubscript{g}. The two regimes correspond to transport residing in the whole InAs transport channel (lower negative $V$\textsubscript{g}) and at the InAs/substrate interface (more negative $V$\textsubscript{g}). \textbf{c}, Conductance as a function of $V$\textsubscript{g} for InAs NWs grown on the GaAs(Sb) buffer along the [110], [1$\bar{1}$0] and [100] directions. \textbf{d}, Resistance, $R$, as a function of $V$\textsubscript{g} for multiple NW samples grown directly on InP, GaAs and the buffer layer.}
 \label{fig4}
 \end{figure}

The lower right insets of Fig. \ref{fig3}\textbf{a-c} show the geometric phase analysis (GPA) of the rotational displacement of the (111) crystal planes. The rotation is a sign of a gradual and partially elastic change of the lattice spacing from GaAs to InAs. An average value for rotation in these structures is on the order of 3$^{\circ}$ (left side) to -3$^{\circ}$ (right side). This signature of elastic strain relaxation is illustrated in a qualitative simulation of the rotational displacement field of a fully strained and partially strained GaAs/InAs NW in Fig. \ref{fig3}\textbf{d}, see \hyperlink{foo}{Methods} for simulation details. Figure \ref{fig3}\textbf{e} shows an image of the GaAs(Sb)//InAs interface of the [1$\bar{1}$0] NW, where non-periodic misfit dislocations (clearly visible after Fourier filtering) indicate partial plastic strain relaxation. Figure \ref{fig3}\textbf{f} shows a zoom-in on a GaAs(Sb)//InAs dislocation free interface region of the [110] NW, where the rotation of the (111) crystal planes peaks at 4.5$^{\circ}$. Here, the interface is nearly dislocation free except for a small region close to the middle of the NW, which means that the NW is close to be fully elastically strain relaxed in the transverse component. The results of the GPA analysis show that most of the elastic strain is released within $\sim$ 20 nm around the interface, where the lattice constant changes from 5.71 to 6.06 \AA{} with a mean dilatation of 6.1 $\%$, see Supplementary Information S6.

We note that the strain relaxation mechanism is similar to that in axial heterostructures of free standing VLS grown NWs\cite{larsson2006strain} and to the elastic rotation previously observed in InGaN/GaN\cite{yoshida2011evidence} and InAs/InSb.\cite{de2014atomic, de2016twin} 
This very important trend opens new possibilities for engineering of elastically strain relaxed SAG NW structures and clearly shows the potential and importance of growth on the top-facet of SAG buffers. This is also apparent from the fact that the InAs NW on the GaAs(Sb) buffer has fewer misfit dislocations than in the case of direct growth on InP substrate, which has a lower lattice mismatch to InAs, see Supplementary Information S6.

\begin{figure*}[ht!]
\vspace{0.2cm}
\includegraphics[scale=1]{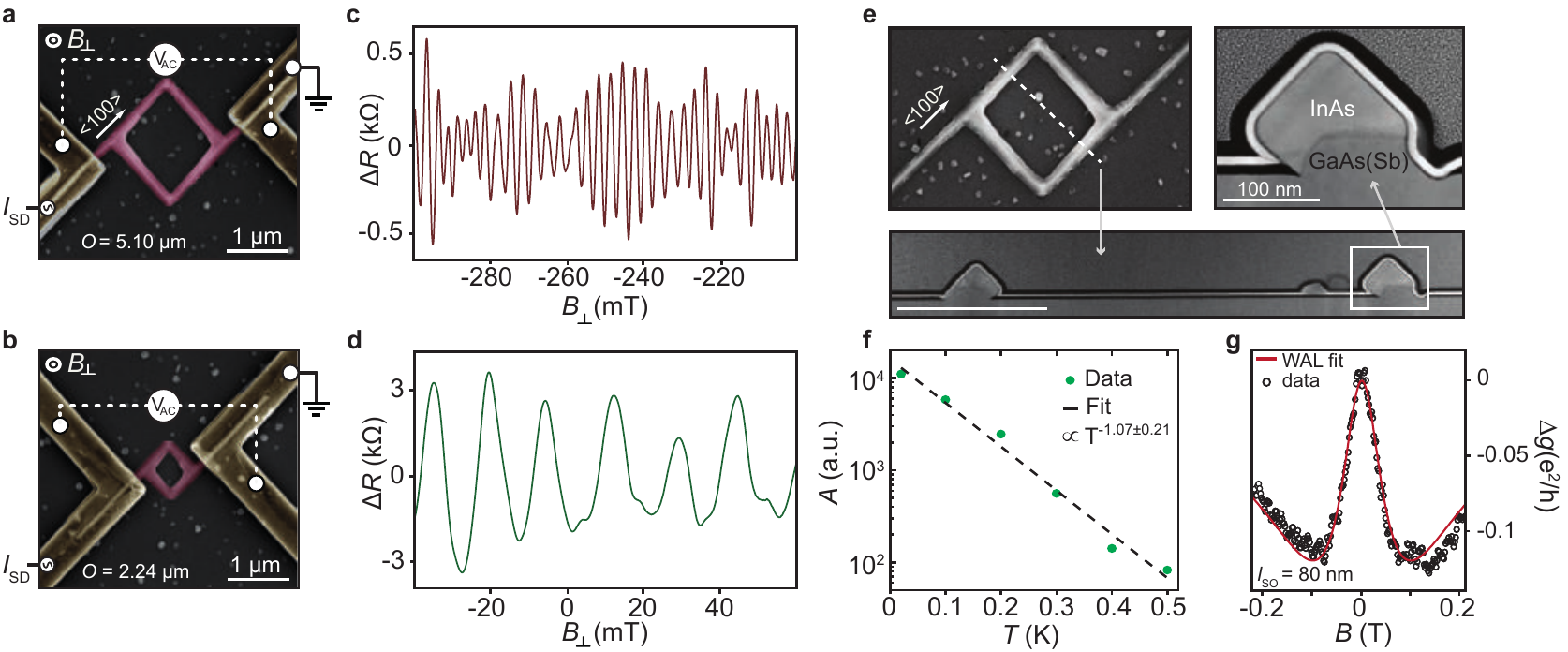}
\vspace{0.2cm}
\caption{\textbf{Low temperature magneto-conductance measurements.} \textbf{a-b}, False-colored SEM micrographs of the four-probe NW loop devices.  \textbf{c-d}, Change in resistance, $\Delta R$, as a function of perpendicular magnetic field, $B_{\perp}$, over a 100 mT range showing electron phase interference oscillations for the larger and smaller loop, respectively. \textbf{e}, Overview SEM and cross-section HAADF-STEM micrographs of the Aharonov-Bohm loop structure showing the asymmetric cross-section. The dashed white line indicates where the cross-section was made. Scale-bar corresponds to 500 nm. \textbf{f}, Integrated FFT of the oscillations plotted as a function of temperature. Fit corresponds to the linear fit of the $h/e$ oscillation where $l_{\phi}$ $\sim$ 13 \textmu m. \textbf{g}, Off-set magneto-conductance traces showing WAL effects around B = 0. The spin-orbit length ($l$\textsubscript{SO} $\sim$ 80 nm) is found from the diffusive regime WAL expression fit.}
\label{fig5}
\end{figure*}

For computing applications \--- classical or quantum \--- the device performance depends on the ability to effectively gate and pinch-off conductance. Since back-gating can be challenging on semi-insulating substrates we use top-gates for the carrier density control. Figure \ref{fig4}\textbf{a} shows a false-colored device lithographically similar to the characterized devices, where only gate width, $W$, is varied. 
Figure \ref{fig4}\textbf{b} shows conductance, $G$, as a function of gate voltage, $V$\textsubscript{g}, for an InAs NW grown on the non-buffered GaAs substrate. The charge carrier density is not fully depleted even for very negative $V$\textsubscript{g} and the down and up gate traces are highly hysteretic, see Supplementary Information S7. This is a general trend for the non-buffered NWs we measured (on both GaAs and InP).
The schematics depict two characteristic gate voltage regions with different slopes of conductance. Due to the top-gate geometry, the carriers in the NW will first be depleted in the top part of the NW, corresponding to the region with the highest slope in Fig. \ref{fig4}\textbf{b}.
As more negative $V$\textsubscript{g} is applied, the carrier density moves towards the NW/substrate interface. The transconductance depends on the quality of the semiconductor crystal. This allows for qualitative distinction between bulk NW and NW/buffer interface properties. Therefore, the interface is mainly probed at more negative gate voltages (as indicated by the sketch in Fig. \ref{fig4}\textbf{b}). For NWs grown directly on the substrate, the bottom interface appears to have a significantly lower field effect mobility. The low interfacial quality caused by the presence of misfit dislocations, potential impurities and roughness originating from the etching process. Additionally, the pre-growth annealing step used to remove the native oxide is likely to play an important role.\cite{van1991oxide}

Figure \ref{fig4}\textbf{c} shows examples of $G$ as a function of $V$\textsubscript{g} for three InAs NWs grown on the buffer layer in the three high-symmetry crystal orientations. The measurements on buffered NWs show a high degree of reproducibility, as all the NWs measured on this growth, pinch-off within a window of $\sim$ 250 mV. In contrast to the non-buffered NWs in Fig. \ref{fig4}\textbf{b}, there is no apparent difference in transconductance between the bulk and the interface gate regions, indicating that the quality of the interfaces is significantly improved.
In Fig. \ref{fig4}\textbf{d} we show data from multiple buffered and the non-buffered NWs of comparable dimensions. Here, the gate dependent resistance is shown for different high-symmetry NW orientations, with $W$ varying from 250 nm to 2.5 $\mu$m. On the logarithmic scale, it is apparent that the conductance is not completely pinched-off in any of the non-buffered NW devices (within reasonable gate range and with selected NW volumes). In addition the non-buffered NWs show more pronounced hysteresis in down/up gate sweeps than the buffered NWs, as presented in Supplementary Information S7. \\

Even though the field effect mobility depends on the gate voltage, $V$\textsubscript{g}, range and carrier density distribution, we use a full fit to the conductance as a function of $V$\textsubscript{g} to extract the mean mobility, as shown in Ref. \citen{gul2015towards}. For details of fitting and finite element modeling of the capacitance see Supplementary Information S8 and S9. The average mean mobility measured on the buffered NWs is about $\bar{\mu}$ $\sim$ 5600 $\pm$ 1300 ${\frac{cm^{2}}{V\cdot s}}$ with a maximum at $\sim$ 7600 ${\frac{cm^{2}}{V\cdot s}}$ (extracted from 24 data sets - 6 buffered devices at 1, 2, 5 and 10 mV bias). More detailed finite element method modeling of the capacitances, including the detailed modeling of the specific cross-sectional shapes, is required for a more exact estimation and for comparison of different NWs and NW orientations. It is clear that the buffer layer significantly improves the transport characteristics of the NWs, regardless of their orientation or cross-sectional area.

We now turn our focus to the quantum transport properties, where quantum phase coherence, scattering length and spin-orbit strength are important characteristics. To study the phase coherence we first fabricated two four-probe loop devices with a circumference of 5.1 \textmu m and 2.24 \textmu m for Aharonov-Bohm (AB) type measurements as shown in Fig. \ref{fig5}\textbf{a} and \textbf{b}, respectively. The resistance shows oscillations in perpendicular magnetic field, $B_{\perp}$, as seen in the 100 mT sweep range in Fig. \ref{fig5}\textbf{c} and \textbf{d}. Magnetic fields on the order of 900 mT were applied along the [110] and [1$\bar{1}$0] in-plane substrate orientations in order to reduce weak anti-localization effects and the aperiodic oscillating background was removed by Savitzky-Golay filtering.\cite{background} The oscillation periods of $\sim$ 2.5 mT (large loop) and $\sim$ 14 mT (small loop) are in good agreement with the areas of the loops, given as $\sim h/(e\cdot area)$. In Fig. \ref{fig5}\textbf{e} we show a cross-sectional TEM image of the loop, as indicated with the dashed line on the SEM image. The asymmetric shape of the InAs NWs on the GaAs(Sb) buffer is related to the growth mechanism of minimizing surface to volume ratio at the junction between the two high-symmetry orientations. We expect this effect to be suppressed with shorter growth time, in line with the discussion on the growth evolution above.

The phase coherence length $l_{\phi}(T)$ can be extracted by fitting the temperature dependence of the AB oscillation amplitude, $A$, obtained from the Fourier spectra (see Supplementary Information S10). Assuming that the amplitude scales as $A(T) \propto exp(-\frac{O}{l_{\phi}(T)})$, where $O$ is the loop circumference\cite{milliken1987effect}, then the exponent $m$ in $l_{\phi}(T)$ $\propto$ T$^{-m}$ can be determined. 
For the small loop we extract the exponent and the phase coherence length by fitting to the logarithm of the amplitude, and get m = 1.07 $\pm$ 0.21  and $l_{\phi}(20mK)$ = 13 $\pm$ 1 \textmu m, see Fig. \ref{fig5}\textbf{f}. It was not possible to obtain a reliable temperature dependence on the large loop due to charge noise switching. 
In the diffusive regime the temperature dependence of the coherence length in a loop shaped structure follows $l_{\phi}$ $\propto$ T$^{-1/2}$, while in the ballistic case with a weak coupling to the environment $l_{\phi}$ $\propto$ T$^{-1}$ as reported in Ref. \citen{hansen2001mesoscopic,Dufouleur2013,seelig2001charge}. This indicates that the small loop resides in the ballistic regime below $\sim$ 500 mK while the non-buffered loop reported in Ref. \citen{sole2018sag} in the diffusive regime.

The dephasing mechanisms are different in a quasi-one dimensional NW and exhibit a different temperature dependence of $l_{\phi}$ $\propto$ T$^{-1/3}$ in the diffusive regime.\cite{ludwig2004interaction,PhysRevB.68.085413,Altshuler1982} With weak anti-localization (WAL) measurements on single NW, we extract a significantly smaller $l_{\phi}$ than compared to the loop measured in Fig. \ref{fig5}\textbf{f}. The theoretical model for the WAL effect, as explained by Ref. \citen{kurdak1992quantum, al1981magnetoresistance}, is fitted to the data. We emphasize that no analytical expression exists for finite 3D cross-section and this type of extraction should only used for comparison between NWs. From the fit we extract a spin-orbit length, $l$\textsubscript{SO}, on the order of 80 nm and $l_{\phi}$ on the order of 180 nm, which is comparable to numbers extracted from similar measurements on VLS grown InAs, InSb and InAs\textsubscript{1-x}Sb\textsubscript{x} NWs.\cite{hernandez2010spin, hansen2005spin,jespersen2018crystal, van2015spin, sestoft2017hybrid}

In conclusion, we show that selective area growth of high-quality InAs NW networks with well defined junctions is feasible in MBE. The NWs can attain significant elastic strain relaxation when grown on top of flat selective area grown buffer layers, with significant improvement of the transport properties in terms of field effect response. Moreover, the material possesses promising quantum transport properties, e.g. strong spin-orbit coupling extracted from WAL and phase coherence demonstrated by AB experiments. We believe that these findings, combined with superconductor epitaxy,\cite{sole2018sag} make this material platform an ideal large-scale architecture for quantum applications that are based on gateable superconducting electronics.




\section{Acknowledgement}
The project was supported by Microsoft Station Q, the European Research Council (ERC) under the grant agreement No.716655 (\textit{HEMs-DAM}), the European Union Horizon 2020 research and innovation program under the Marie Sk\l{}odowska-Curie grant agreement No 722176, the Danish National Science Research Foundation and the Villum Foundation. We thank Chris Palmstr\o m, Philippe Caroff, Lucia Sorba and Roman Lutchyn for fruitful discussions, and thank Claus B. S\o rensen, Robert McNeil, Karthik Jambunathan and Shivendra Upadhyay for technical assistance in Copenhagen. SMS acknowledges funding from "Programa Internacional de Becas "la Caixa"-Severo Ochoa". JA and SMS also acknowledge funding from Generalitat de Catalunya 2017 SGR 327. ICN2 acknowledges support from the Severo Ochoa Programme (MINECO, Grant no. SEV-2013-0295) and is funded by the CERCA Programme / Generalitat de Catalunya. Part of the present work has been performed in the framework of Universitat Aut\`{o}noma de Barcelona Materials Science PhD program. The HAADF-STEM microscopy was conducted in the Laboratorio de Microscopias Avanzadas at the Instituto de Nanociencia de Aragon-Universidad de Zaragoza. JA and SMS thank them for offering access to their instruments and expertise.

\section{Author Contributions}
The crystal growth was done by FK, JES, PA, FB and PK with support from YL, RK and EU, and substrate fabrication by FK, JES, LC and AF. Device fabrication was done by FK, JES, SV, LC. Atomic force microscopy characterization by SAK. Transmission electron microscopy based characterization, sample preparation and related analysis (including GPA and EELS) was carried out by SMS and JA, with strain analysis and simulation by TS, FK, JES, PK. Electronic characterization and analysis was done by JES, FK, SV, LC, AW, LPK, CMM and PK. FK, JES and PK wrote the paper with contributions from all authors. 

\section{Competing financial interests}
The authors declare no competing financial interests.


\hypertarget{foo}{\section{Methods}}

\paragraph{Substrate preparation.} Semi-insulating and epi-ready Fe doped InP and un-doped GaAs substrates were covered with 10-30 nm of SiO$_{x}$ in a SPTS Multiplex plasma enhanced chemical vapor deposition system at 300$^{\circ}$C. The patterning of the wafers was done by standard electron beam lithography processing, and the oxide mask was etched in an ammonium fluoride solution or reactive ion etching before stripping the resist in acetone.

\paragraph{SAG NW growth by MBE.} InP/GaAs substrates with SiO$_{x}$/SiN$_{x}$ masks, were used for non-buffered selective area growth of InAs. The native oxide was removed under an As\textsubscript{2} overpressure following standard procedure on bare substrates. InAs was grown at around 500$^{\circ}$C with a corresponding planar growth rate of $\sim$ 0.1 \textmu m/h for InAs and 0.26 \textmu m/h for GaAs substrates. For buffered growth on GaAs the native oxide was blown-off as described above, before GaAs(Sb) was grown under similar conditions using a corresponding planar growth rate of $\sim$ 0.1 \textmu m/h and an As\textsubscript{2}/Sb\textsubscript{2+4} ratio of 7. Following InAs growth on the buffered GaAs(Sb) on GaAs was carried out at the same temperature with a corresponding planar growth rate of $\sim$ 0.1 \textmu m/h.

\paragraph{Cross-section STEM characterization.}  
Selected devices were covered with 15 nm of ALD Al\textsubscript{2}O\textsubscript{3} and thinned into cross-sectional lamellae. Several cross-sectional lamellae on selected NWs were cut by using a focused ion beam. TEM Characterization: HR-TEM, HAADF scanning TEM (STEM) and EELS spectrum imaging (SI) were performed using a TECNAI F20 field emission gun microscope operated at 200 kV with a point to point resolution of 0.14 nm coupled to a GATAN Quantum EELS spectrometer. The atomic resolution HAADF-STEM images were acquired on a probe corrected FEI Titan 60-300 equipped with a high brightness field emission gun (XFEG) and a CETCOR corrector from CEOS to produce a probe size below 1 \AA. The microscope was operated at 300 kV, with a convergence angle of 25 mrad and an inner collection angle of the detector of 58 mrad. Atomic resolution aberration corrected HAADF-STEM was used to determine with high accuracy the atomic column positions, allowing the detailed study of polarity \cite{de2012polarity} as well as the final strain analysis by means of GPA \cite{hytch1998quantitative}.

\paragraph{Elastic strain simulations.}
Rotational deformation fields were simulated using finite element method software COMSOL Multiphysics$^{\textregistered}$. A 2D model was made, consisting of GaAs and InAs parts with a continuous boundary in between. Initial compressive strain was applied to InAs, in order to simulate the effect of the lattice mismatch. A stable solution of elastic strain distribution was then found assuming anisotropic linear elasticity, with elastic constants for GaAs\cite{cottam1973elastic} and InAs\cite{shur1996handbook}. Presence of dislocations at the interface results in lower values of elastic relaxation in a form of lattice rotation, compared to the fully epitaxial case.

\paragraph{Device fabrication.} Devices were fabricated directly on the growth substrate by spinning poly-MMA resist at 4000 RPM for 45 s and baked at 185$^{\circ}$C. Standard EBL procedures were used to expose contacts. The native oxide was removed by RF Argon ion milling at 15 W for 5 min before depositing 10 nm Ti and 180 nm Au as ohmic contacts. 8 nm of ALD HfO\textsubscript{2} was grown at 90$^{\circ}$C, before standard EBL patterning followed by deposition of Au top-gates.

\paragraph{5 K measurements.}
The samples were cooled and out-gassed overnight in a Lakeshore 4K cryo-free probestation operating at a base temperature of T $\sim$ 5 K. Standard four-probe measurements were performed using a Keithley 2600 sourcemeter to control the gate voltage, the source-drain bias and measure the current. An Agilent digital multimeter was used to probe the voltage drop on the inner contacts. 

\paragraph{20 mK measurements.}

The WAL measurements were carried out using standard current-biased ac lock-in techniques in a Bluefors XLD-400 dilution refrigerator operating at a base temperature of T $\sim$ 20 mK. 
The AB experiments were carried out using current-biased ac lock-in technique in a Oxford Triton 100 dilution refrigerator with a base temperature of T $\sim$ 20 mK. Magnetic fields aligned to the [110] and [1$\bar{1}$0] directions on the order of 900 mT were applied to suppress the WAL effect during the measurement while sweeping the perpendicular magnetic field.

\bibliography{ref}

\end{document}


\title{Supplementary Information:\\ \vspace{0.2cm} Field effect enhancement in buffered quantum nanowire networks}

\author{Filip Krizek\textsuperscript{\textdagger}}
\affiliation{Center For Quantum Devices and Station Q Copenhagen, Niels Bohr Institute, University of Copenhagen, 2100 Copenhagen, Denmark}
\altaffiliation{These authors contributed equally to this work.}

\author{Joachim E. Sestoft\textsuperscript{\textdagger}}
\affiliation{Center For Quantum Devices and Station Q Copenhagen, Niels Bohr Institute, University of Copenhagen, 2100 Copenhagen, Denmark}
\altaffiliation{These authors contributed equally to this work.}

\author{\mbox{Pavel Aseev\textsuperscript{\textdagger}}}
\affiliation{QuTech and Kavli Institute of Nanoscience, Delft University of Technology, 2600 GA Delft, The Netherlands}
\altaffiliation{These authors contributed equally to this work.}

\author{Sara Marti-Sanchez}
\affiliation{Catalan Institute of Nanoscience and Nanotechnology (ICN2), CSIC and BIST, Campus UAB, Bellaterra, Barcelona, Catalonia, Spain}

\author{Saulius Vaitiek\.{e}nas}
\affiliation{Center For Quantum Devices and Station Q Copenhagen, Niels Bohr Institute, University of Copenhagen, 2100 Copenhagen, Denmark}

\author{Lucas Casparis}
\affiliation{Center For Quantum Devices and Station Q Copenhagen, Niels Bohr Institute, University of Copenhagen, 2100 Copenhagen, Denmark}

\author{Sabbir A. Khan}
\affiliation{Center For Quantum Devices and Station Q Copenhagen, Niels Bohr Institute, University of Copenhagen, 2100 Copenhagen, Denmark}

\author{Yu Liu}
\affiliation{Center For Quantum Devices and Station Q Copenhagen, Niels Bohr Institute, University of Copenhagen, 2100 Copenhagen, Denmark}

\author{Toma\v{s} Stankevi\v{c}}
\affiliation{Center For Quantum Devices and Station Q Copenhagen, Niels Bohr Institute, University of Copenhagen, 2100 Copenhagen, Denmark}

\author{Alexander M. Whiticar}
\affiliation{Center For Quantum Devices and Station Q Copenhagen, Niels Bohr Institute, University of Copenhagen, 2100 Copenhagen, Denmark}

\author{Alexandra Fursina}
\affiliation{Microsoft Station Q, Delft University of Technology, 2600 GA Delft, The Netherlands}

\author{Frenk Boekhout}
\affiliation{QuTech and Netherlands Organization for Applied Scientific Research (TNO), Stieltjesweg 1, 2628 CK Delft, The Netherlands}

\author{Rene Koops}
\affiliation{QuTech and Netherlands Organization for Applied Scientific Research (TNO), Stieltjesweg 1, 2628 CK Delft, The Netherlands}

\author{Emanuele Uccelli}
\affiliation{QuTech and Netherlands Organization for Applied Scientific Research (TNO), Stieltjesweg 1, 2628 CK Delft, The Netherlands}

\author{Leo P. Kouwenhoven}
\affiliation{QuTech and Kavli Institute of Nanoscience, Delft University of Technology, 2600 GA Delft, The Netherlands}
\affiliation{Microsoft Station Q, Delft University of Technology, 2600 GA Delft, The Netherlands}

\author{Charles M. Marcus}
\affiliation{Center For Quantum Devices and Station Q Copenhagen, Niels Bohr Institute, University of Copenhagen, 2100 Copenhagen, Denmark}

\author{Jordi Arbiol}
\affiliation{Catalan Institute of Nanoscience and Nanotechnology (ICN2), CSIC and BIST, Campus UAB, Bellaterra, Barcelona, Catalonia, Spain}
\affiliation{ICREA, Pg. Llu\'{i}s Companys 23, 08010 Barcelona, Catalonia, Spain}

\author{Peter Krogstrup}
\email{krogstrup@nbi.dk}
\affiliation{Center For Quantum Devices and Station Q Copenhagen, Niels Bohr Institute, University of Copenhagen, 2100 Copenhagen, Denmark}

\date{\today}

\maketitle

\subsection{S1. Al deposition} 
\begin{figure}[htb!]
\vspace{0.2cm}
\includegraphics[scale=0.2]{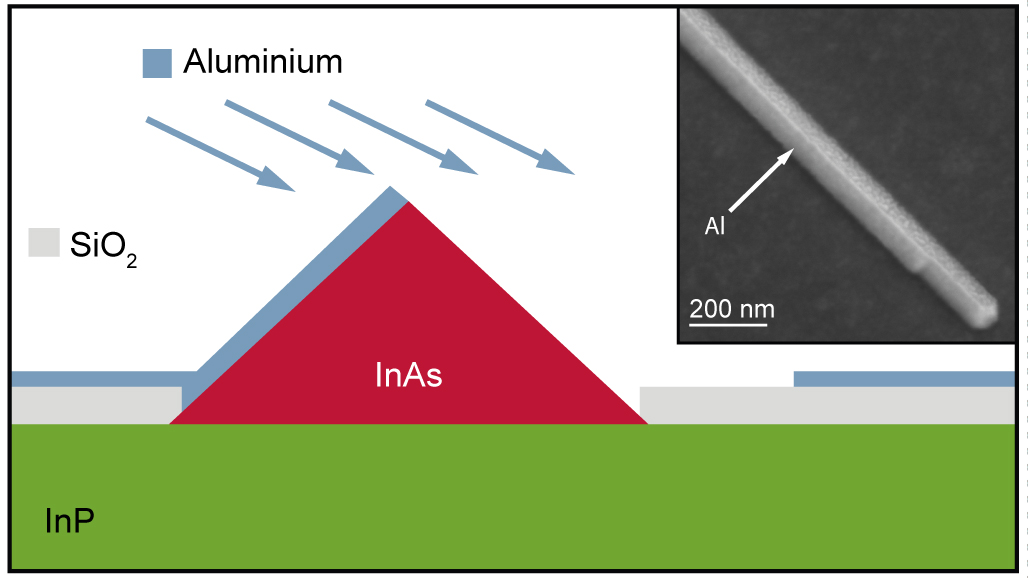}
\vspace{0.2cm}
\caption{\textbf{Al deposition on one facet of a [100] NW.} Schematic showing the directional Al deposition onto one facet of the NW. Inset shows an SEM micrograph of InAs NW grown on InP substrate with directionally deposited Al. }
\label{fig1}
\end{figure}

The fully developed cross-sectional equilibrium shape of the NWs allows for directional deposition of a superconductor (or other materials) on selected facets of the NWs, see illustration in Fig. \ref{fig1}. In hybrid semiconductor-superconductor devices, this is an advantageous geometry as it allows for tuning of the carrier density with side/top gates. Full shell devices are difficult to gate with standard side-gate geometries, due to the superconductor electrostatic screening. In the inset of Fig. \ref{fig1} we show an InAs NW with Al deposited from an angle of 33$^\circ$ from the wafer normal, this resulted in a uniformly coated facet, but in this particular case for [100] orientated NWs, it also results in deposition of an undesirable thin layer on the second NW facet. Regardless, the density in the NW could still be tuned with a side gate, because the thin dewetted layer oxidizes as it is taken out of the growth chamber. 

\subsection{S2. Substrate fabrication} 


Semi-insulating Fe-doped InP and undoped GaAs (001) substrates are utilized because of their compatibility with as-grown device fabrication and measurements at low temperatures. First 10-30 nm of SiO\textsubscript{x} or SiN\textsubscript{x} is grown by plasma-enhanced chemical vapour deposition (PECVD) on the epi-ready wafers. Mask openings were defined by standard electron beam lithography (EBL). For the SiO\textsubscript{x} mask a short acid dip (3 s in buffered hydro-fluoride, BHF) is used to strip the oxide in the defined polymer windows. For SiN\textsubscript{x} masks reactive ion etching was used to create the mask opening.

\begin{figure}[b!]
\vspace{0.2cm}
\includegraphics[scale=0.25]{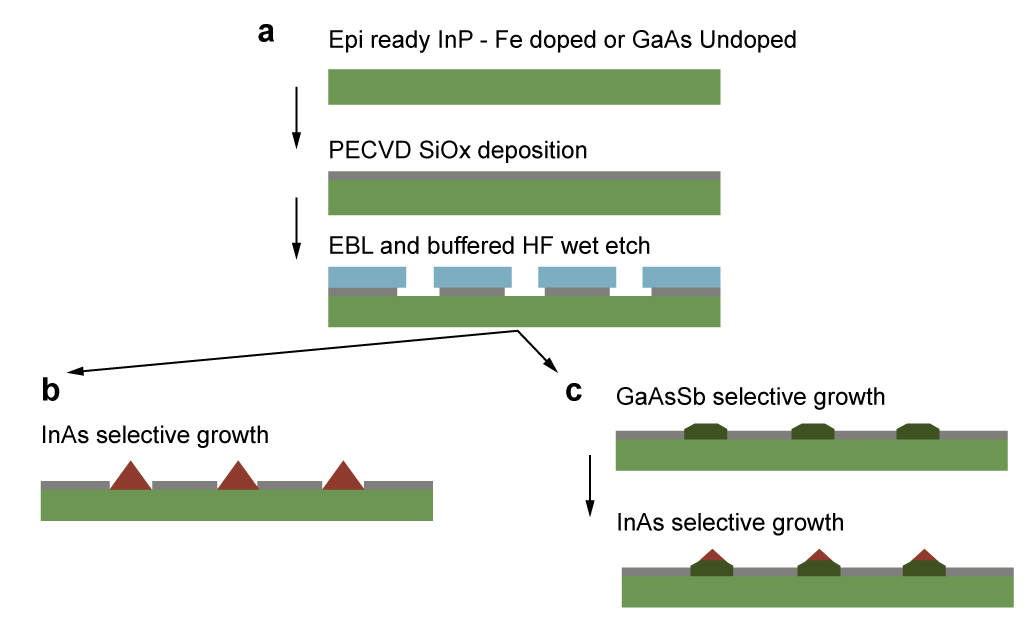}
\vspace{0.2cm}
\caption{\textbf{Substrate fabrication and material growth.} \textbf{a}, The growth wafer (InAs or GaAs) is illustrated in green, the dielectric mask (SiOx or SiN)  is indicated in grey, whereas PMMA is colored in blue.. Grey is the dielectrics SiO\textsubscript{x} or SiN\textsubscript{x}, blue is PMMA. The mask is defined by standard EBL processes and the dielectric layer is selectively etched. After resist lift-off the substrates are desorbed at 250 $^{\circ}$C  and \textbf{b}, InAs or \textbf{c}, GaAs(Sb) and then InAs is grown by MBE. Red is InAs and dark green is the buffer, GaAs(Sb).}
\label{fig2}
\end{figure}

After thorough cleaning in ultra-pure milli-q water, and iso-propanol and acetone, the samples are subjected to O$_{2}$ plasma ashing before the wafer is loaded into the MBE system and degassed in UHV at 250 $^{\circ}$C (for InP) or 350 $^{\circ}$C (for GaAs) for two hours before entering the growth system. See Fig. \ref{fig2} for illustration of fabrication principle.

\vspace{-0.2cm}
\subsection{S3. Non-buffered InAs SAG on GaAs (001) substrates}
\vspace{-0.5cm}
\vspace{-0.2cm}
\begin{figure}[ht!]
\vspace{0.15cm}
\includegraphics[scale=0.79]{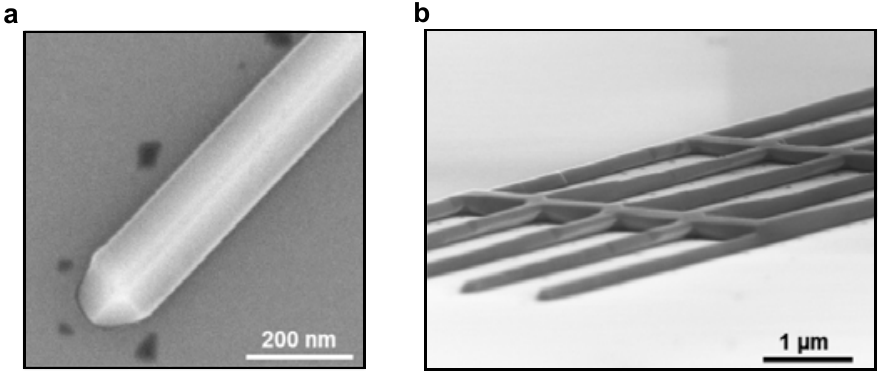}
\vspace{-0.2cm}
\caption{\textbf{Un-buffered InAs NWs on GaAs(001) substrate.} \textbf{a}, SEM micrograph of a single InAs NW on GaAs (001). \textbf{b}, InAs NWs network on GaAs (001) with double the growth time compared to \textbf{a}.}
\label{fig3}
\end{figure}
\vspace{-0.4cm}


\subsection{S4. AFM of [110]/[1-10] junction}
\begin{figure}[ht!]
\vspace{0.15cm}
\includegraphics[scale=0.25]{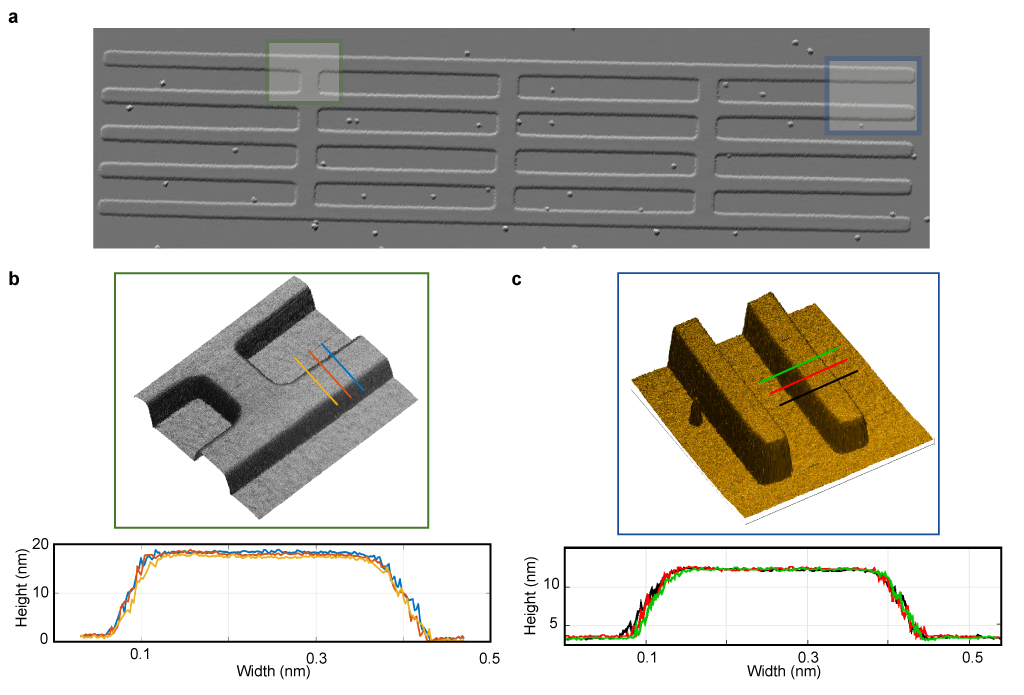}
\vspace{-0.8cm}
\caption{\textbf{AFM of the GaAs(Sb) buffer.}~\textbf{a}, AFM micrograph of [110]/[1-10] GaAs(Sb) buffer layer grown in a network shape. \textbf{b}, 3-dimensional AFM image of the junction indicated by the green box in \textbf{b}. Three line traces measured across one of the legs of the junction. \textbf{c}, Same as \textbf{c}, but corresponding to the blue box in \textbf{a}.}
\label{fig4}
\end{figure}

Here we investigate the flatness of the [100] top facet of the GaAs(Sb) buffer in order to see whether the orientation of the grown buffer has an impact on the flatness. The GaAs(Sb) buffer layer was investigated by atomic force microscopy (AFM) imaging of the [110]/[1-10] networks (as opposed to the [100]/[010] junctions in the main text). We find that the roughness of the [110]/[1-10] junction is comparatively similar to the [100]/[010] junction roughness, as shown in Fig. \ref{fig4}. 

\vspace{-1cm}

\newpage

\subsection{S5. InAs nanowire grown on top facet - with width smaller than the mask width}

\begin{figure}[ht!]
\vspace{0.2cm}
\includegraphics[scale=0.12]{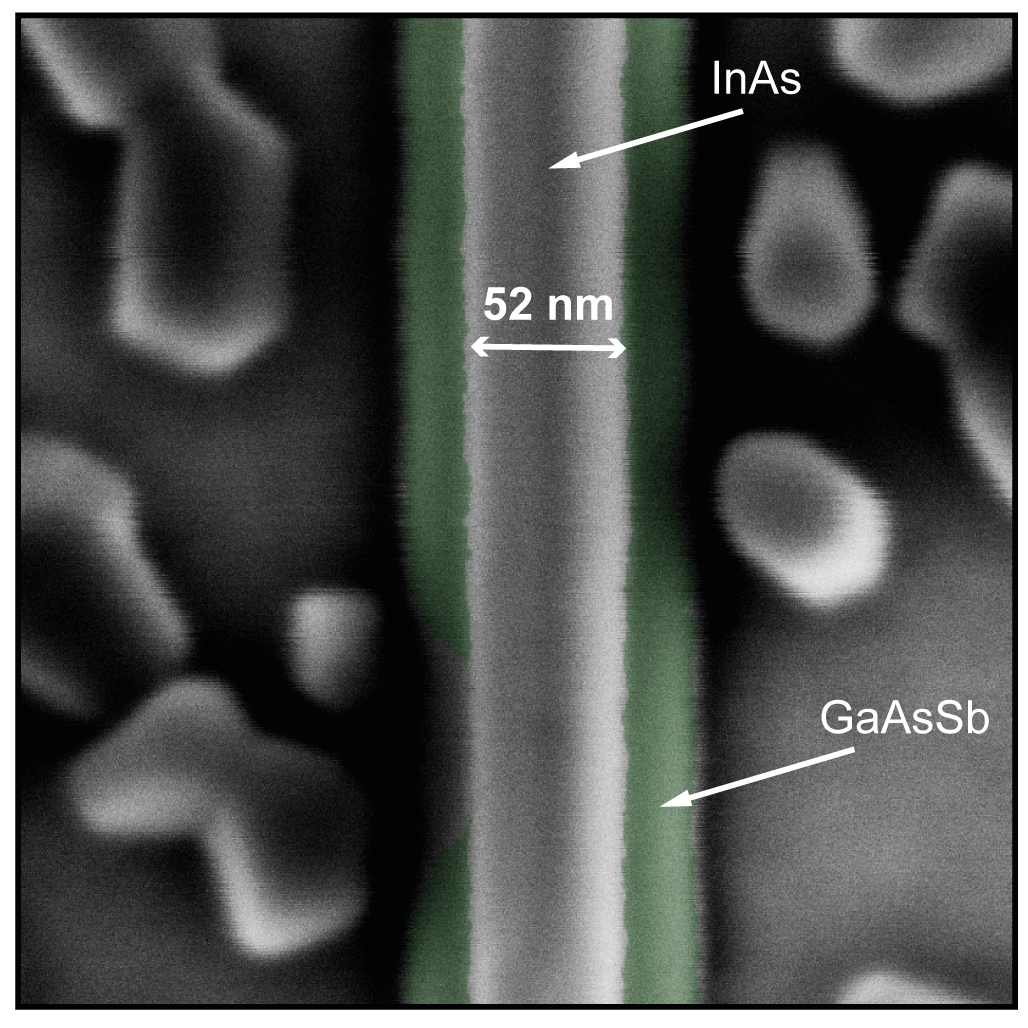}
\vspace{0.2cm}
\caption{\textbf{Thin InAs nanowires.} SEM micrograph of InAs NW grown on GaAs(Sb) buffer layer.}
\label{fig5}
\end{figure}

One of the advantages of selectively growing NWs on a buffer layer is that it allows for reducing the NW diameter below the width of the mask opening, as shown in Fig. \ref{fig5}. 

Growth solely on the top facet is seen for thin layers, as also reported on GaSb buffers in Ref. [\citen{fahed2016selective}]. Additionally, this approach is advantageous because the transport channel is not in direct contact with neither the processed substrate surface nor the oxide mask, especially since fabrication of mask openings with atomically smooth sidewalls is challenging.   
\begin{figure}[ht!]
\vspace{0.2cm}
\includegraphics[scale=1.0]{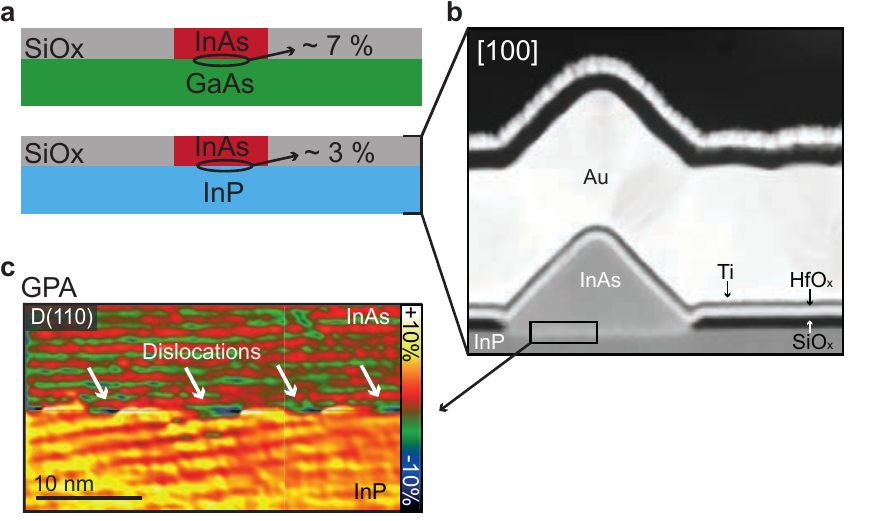}
\vspace{-0.2cm}
\caption{\textbf{InAs/InP interface.} \textbf{a}, The strain of InAs to GaAs is 7 \%, and 3 \% between InAs and InP. \textbf{b}, High resolution aberration corrected HAADF-STEM image of the InAs and InP interface. The cross-section is prepared by FIB processing over a gate, explaining the layer sequence. \textbf{a}, Geometrical phase analysis of the InAs/InP interface shows a fully strain relaxed interface by periodic misfit dislocations. }
\label{fig6}
\end{figure}

\begin{figure*}[t!]
\vspace{0.2cm}
\includegraphics[scale=0.4]{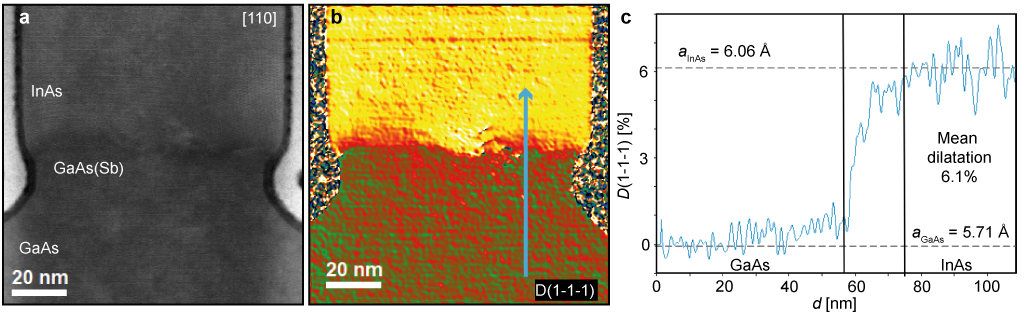}
\vspace{0.2cm}
\caption{\textbf{InAs/GaAs(Sb) interface.} \textbf{a}, HAADF-STEM micrograph of a [110] oriented NW. \textbf{b}, GPA dilatation map of the same NW taken along the (1-1-1) planes. \textbf{c}, Plot of dilatation, $D$, as a function of vertical distance, $d$, along the cut through the NW interface.}
\label{fig7}
\end{figure*}

\subsection{S6. Non-buffered interface to substrate}

The NWs were grown directly on GaAs (7 \% strain) and InP (3 \% strain) substrates, as sketched in Fig. \ref{fig6} \textbf{a}. The devices used for transport measurements in the main text are sliced into thin lamellae using a focused ion beam (FIB) on the top gate structures, for more details see Methods. The samples are investigated by atomic-resolution aberration-corrected high-angle-annular-dark-field scanning transmission electron microscopy (STEM) as seen in Fig. \ref{fig6} \textbf{b}.

The HfO\textsubscript{x} is used as a dielectric before evaporating Ti and Au to form the top gate. On top of the whole structure a thick layer of Pt is deposited to protect the structure from Ga implementation and amorphization during the FIB process.

Geometrical phase analysis (GPA) is performed on the interface between the InAs and InP where the two structure are fully in-elastically relaxed at the interface, as indicated by the presence of misfit dislocations running parallel to the interface in Fig. \ref{fig6} \textbf{c}. The GPA of the dilatation of the [110] oriented NW in Fig. \ref{fig7} \textbf{a} and \textbf{b} shows that the lattice constant fully relaxes from GaAs to InAs across the $\sim$ 20 nm buffer-region.

\subsection{S7. Hysteresis on buffered and non-buffered NW devices}

\begin{figure}[ht!]
\vspace{0.2cm}
\includegraphics[scale=1]{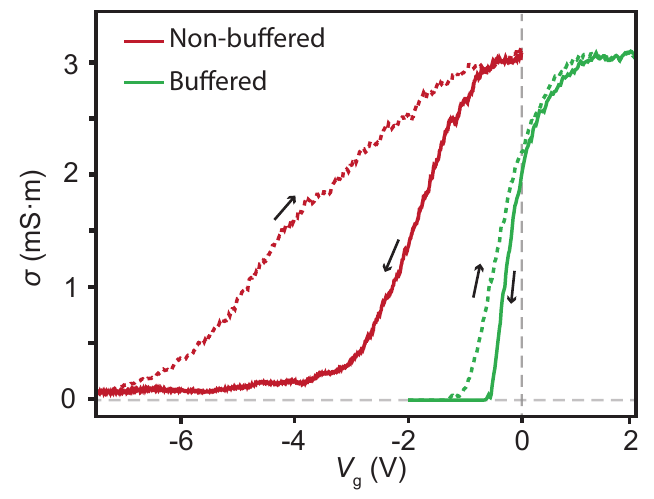}
\vspace{0.2cm}
\caption{\textbf{Comparison of NW hysteresis.} Up and down sweeps of the four-probe gate traces for non-buffered and buffered NW devices measured at the temperature of $\sim$ 5 K. }
\label{fig8}
\end{figure}

In Fig. \ref{fig8} we show examples of conductivity, $G$, traces as a function of the gate voltage, $V$\textsubscript{g}, for a non-buffered and a buffered device. Here the “non-buffered” refers to an InAs NW grown directly on InP substrate while the buffered to InAs NW grown on a GaAs(Sb) buffer on a GaAs substrate. The growth of the InAs transport channel on top of the GaAs(Sb) buffer layer significantly improves the carrier density tuning and hysteresis. This improvement was apparent regardless of the NW orientation or dimensions.

\subsection{S8. Finite element method capacitance modeling}
\begin{figure*}[ht!]
\vspace{0.2cm}
\includegraphics[scale=0.4]{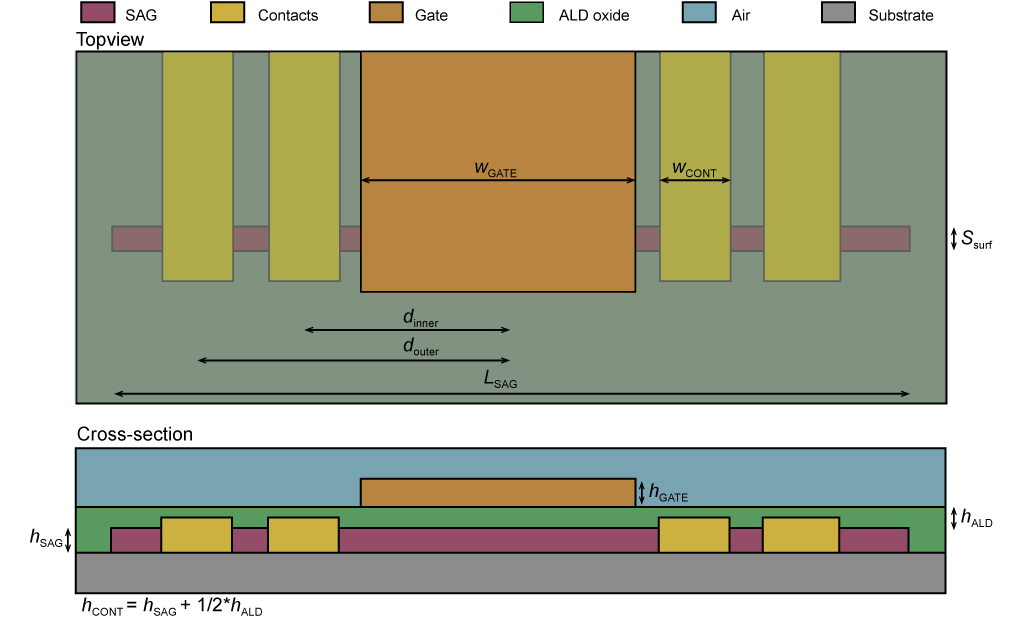}
\vspace{0.2cm}
\caption{\textbf{Capacitance model schematics.} Schematics of the geometry used in the finite element method simulation, containing all the geometrical input parameters. }
\label{fig9}
\end{figure*}

\begin{figure*}[ht!]
\vspace{0.2cm}
\includegraphics[scale=0.4]{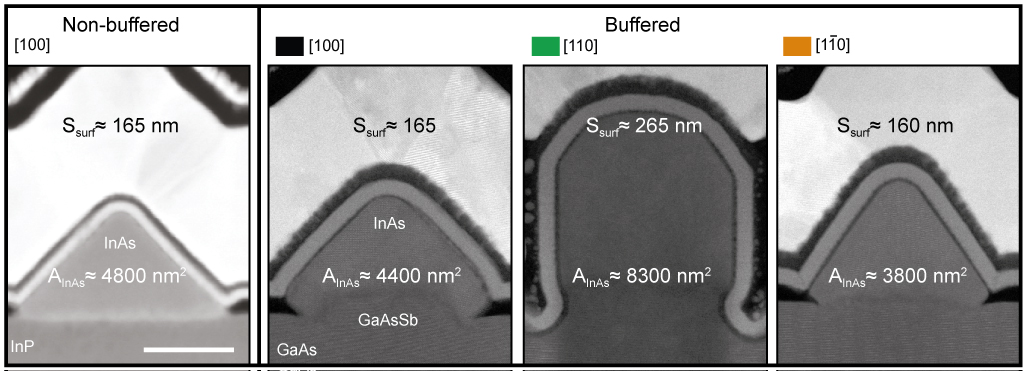}
\vspace{0.2cm}
\caption{\textbf{Nanowire device dimensions.}
Cross-sectional TEM micrographs of non-buffered and buffered InAs NWs, highlighting the cross-sectional area of the InAs transport channel AInAs and the NW/gate surface $S$\textsubscript{surf}.}
\label{fig10}
\end{figure*}

The device capacitances were simulated using the finite element method in the electrostatic module in COMSOL Multiphysics. Figure \ref{fig9} summarizes the geometrical input parameters of the model, where the exact values were extracted from SEM images of the devices and geometrical cross-sections as shown in Fig. \ref{fig10}. The geometry of the NW was approximated by a rectangular cross-section with the width corresponding to $S$\textsubscript{surf} as also shown in Fig. \ref{fig9}.

Two additional approximations were that the height of the gold contacts is set as $h$\textsubscript{CONT} = $h$\textsubscript{SAG}$+1/2\cdot$ $h$\textsubscript{ALD} and that the real height of the atomic layer deposition (ALD) oxide layer $h$\textsubscript{ALD} corresponds to reality between the NW and the top-gate, but elsewhere it is given as $h$ = $h$\textsubscript{SAG}$+h$\textsubscript{ALD}. Another parameter is the effective dielectric constant of the used HfO\textsubscript{x} in combination with the NW native oxide, which was estimated as $k$ = 6.5. 

\subsection{S9. Field effect mobility fit}
\begin{figure}[ht!]
\vspace{0.2cm}
\includegraphics[scale=1.2]{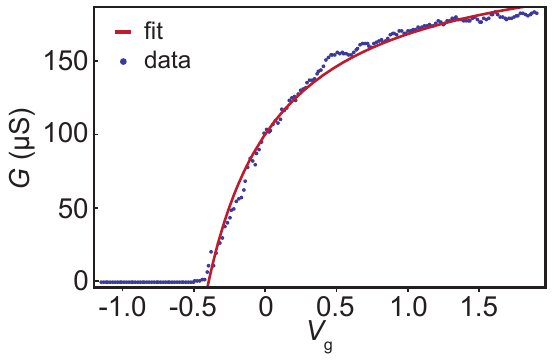}
\vspace{0.2cm}
\caption{\textbf{Field-effect mobility fit to pinch-off curve.} Data is conductance, $G$, as a function of gate voltage, $V$\textsubscript{g} shown as blue dots. The fit is the red line.}
\label{fig12}
\end{figure}

The field effect mobility is extracted for the buffered devices by fitting the equation from Ref. [\citen{gul2015towards}], 

$$ G(V_g) = 1 /\Big( R_s+\frac{W^2}{\mu C(V_g-V_{th})} \Big),$$ 

to the pinch-off curves measured at temperatures on the order of 5 K, as shown in Fig. \ref{fig12}. Here $G(V_g)$ is the conductance, $W$ is the length of the InAs channel and $\mu$, $V_{th}$ and probe resistance $R_{s}$ are the fit parameters. The $C$ is found from the finite element modeling described in S8.

\subsection{S10. Aharonov Bohm oscillations and frequency extraction
}
The Fourier transform of the data in Fig. 5 from the main text, showing the normalized amplitude of the $h$/$ne$ oscillations. The largest amplitude denoted, $A_{1}$, corresponds to the $h$/$ne$ oscillation. The higher frequency peaks, corresponding to $A_{2}$ and $A_{3}$, could be attributed to two things: (1) The asymmetric cross-section of the InAs NWs grown on the GaAs(Sb) buffer as seen in Fig. 5 in the main text, results in more than one effective circumference and thus contain multiple available transport paths. (2) Electronic waves propagating around the loop more than once results
in higher order harmonics. The two dashed lines corresponding to $A$\textsubscript{out} and $A$\textsubscript{in} mark the outer and inner areas of the loop. Since the higher harmonics are not residing at double the frequency of $A_1$ and all amplitudes are within the area bound $A$\textsubscript{out} and $A$\textsubscript{in} explanation (1) is more probable. 

\begin{figure}[ht!]
\vspace{0.2cm}
\includegraphics[scale=1.2]{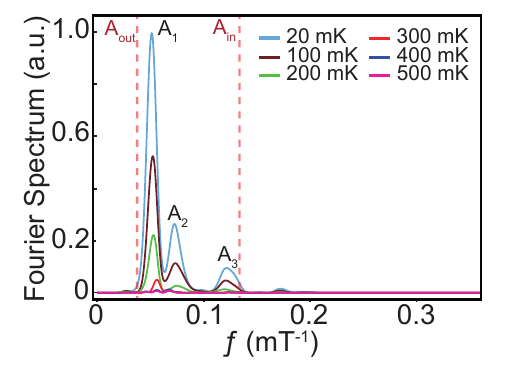}
\vspace{0.2cm}
\caption{\textbf{Integrated FFT of the AB oscillations.}
Normalized Fourier spectrum amplitude of the Aharonov Bohm oscillations as a function of frequency at temperature range varying from 20 to 500 mK. }
\label{fig13}
\end{figure}

\bibliography{ref}